\documentstyle[12pt]{article}
\advance\textheight by \topskip
\begin{document}
\thispagestyle{empty}
\begin{flushright}
SU--ITP--93--36\\
IEM--FT--82/93\\
astro-ph/9401042\\
January 24, 1994
\end{flushright}
\vskip 2cm
\begin{center}
{\Large\bf Jordan--Brans--Dicke Stochastic Inflation}\\
\vskip 1.5cm
{\bf Juan Garc\'{\i}a--Bellido}\footnote{
E-mail: bellido@slacvm.slac.stanford.edu}
\vskip 0.05cm
Department of Physics, Stanford University, \\
Stanford, CA 94305-4060, USA
\end{center}

\vskip 2cm

{\centerline{\large\bf Abstract}}
\begin{quotation}
\vskip -0.4cm
We study stochastic inflation in the presence of a dynamical
gravitational constant. We describe the Arnowitt--Deser--Misner
formalism for Jordan--Brans--Dicke theory of gravity with an
inflaton field. The inflaton and dilaton scalar fields can be
separated into coarse-grained background fields and quantum
fluctuations. We compute the amplitude of the perturbations
generated by those quantum fluctuations in JBD theory with an
arbitrary potential for the inflaton field. The effect of the
quantum fluctuations on the background fields is equivalent to a
Brownian motion of the scalar fields, which can be described
with the use of a Fokker--Planck diffusion equation. The
probability to find a given value of the fields in the comoving
frame can be written as a Gaussian distribution centered on
their classical trajectory, with decreasing dispersion along
both field directions. We also calculate the condition for the
Universe to enter a self-regenerating inflationary phase.  The
probability distribution in the physical frame, which takes into
account the expansion of the proper volume of the inflationary
domains, will be concentrated at the Planck boundary and will
move along it towards large values of the fields.
\end{quotation}
\newpage

\def\lsim{\mathrel{\lower2.5pt\vbox{\lineskip=0pt\baselineskip=0pt
           \hbox{$<$}\hbox{$\sim$}}}}
\def\gsim{\mathrel{\lower2.5pt\vbox{\lineskip=0pt\baselineskip=0pt
           \hbox{$>$}\hbox{$\sim$}}}}
\def\eel#1{\label{#1}\end{equation}}
\newcommand{\be}{\begin{equation}}
\newcommand{\ba}{\begin{array}}
\newcommand{\ee}{\end{equation}}
\newcommand{\ea}{\end{array}}
\newcommand{\form}[1]{(\ref{#1})}
\newcommand{\<}{\left\langle}
\renewcommand{\>}{\right\rangle}
\newcommand{\med}{\frac{1}{2}}
\newcommand{\sig}{\sigma}
\newcommand{\lam}{\lambda}
\newcommand{\eps}{\epsilon}
\newcommand{\w}{\omega}
\newcommand{\T}{T_{\rm bh}}
\newcommand{\M}{M_{\rm p}}
\newcommand{\bM}{\bar{M}_{\rm p}}
\newcommand{\spl}{s_{\rm p}}
\newcommand{\hs}{\hspace{5mm}}
\newcommand{\vs}{\vspace{2mm}}
\newcommand{\vx}{\vec{x}}
\newcommand{\vk}{\vec{k}}

\section{\label{Intro} Introduction}

The early inflationary scenarios could not solve but only
postpone the problem of initial conditions for the Big Bang. The
first models \cite{InfCos} assumed that the universe started in
a very hot state that supercooled in a metastable vacuum, which
then decayed to the true vacuum through a first order phase
transition or just rolled down through a second order phase
transition. Linde's chaotic inflation \cite{InfCos} opened the
possibility of arbitrary initial conditions for arbitrary
effective potentials for the inflaton field. It was soon
realized \cite{LinBook} that the most natural initial conditions
for inflation were close to Planck scale, where quantum
fluctuations of the space-time metric become important.

With the inclusion of the fields' quantum fluctuations, it was
understood \cite{Eternal,SelfRep} that the whole Universe might
be in eternal self-reproduction of inflationary domains, while
our own observable universe could be just a late subproduct of
one of such domains. The Brownian motion of the inflaton field
under the effect of its own quantum fluctuations can be
described with the help of diffusion equations, in the context
of the so-called stochastic inflation formalism
\cite{StoInf,SelfRep}.  This formalism describes the interplay
between quantum fluctuations and coarse-grained background
fields, and allows one to study the very large scale structure
of the Universe \cite{GonLin}--\cite{LLM}, much beyond our
observable universe.  It is also a consistent framework for the
description of the origin of density perturbations during
inflation, that may have given rise to the recently observed
temperature fluctuations in the cosmic background radiation
\cite{COBE} and, furthermore, to the density inhomogeneities for
galaxy formation.

Stochastic inflation stresses the fact that inflation is a
random process, in which quantum fluctuations of the fields act
as stochastic forces on their own background values as a result
of the expansion in de Sitter space. The universe is divided
into causally independent inflationary domains, in which the
fields acquire different values. The global description of this
stochastic process uses diffusion equations for the probability
distribution to find a given value of the fields in a given
inflationary domain. The probabilistic description of stochastic
inflation opened the possibility of a connection with quantum
cosmology \cite{LinBook,Mijic,LLM}, in the sense that it may
provide an interpretation of the wave function of the universe
as the probability amplitude for an ensemble of inflationary
domains, instead as for the whole Universe.

It is generally assumed in stochastic inflation that the
gravitational dynamics of the whole Universe is correctly
described by general relativity. However, it seems a strong
assumption to extrapolate the description of the gravitational
phenomena at our local and low energy scales to the very large
scales beyond our observable universe. In fact, it is believed
that the theory of general relativity is just a low energy
effective theory of the gravitational interaction at the quantum
level.  So far the only consistent but by no means definite,
since we lack the experimental observations needed to confirm
it, theory of quantum gravity is string theory \cite{GSW}.
String theory contains in its massless gravitational sector a
dilaton scalar field as well as the graviton. The low energy
effective theory from strings has the form of a scalar-tensor
theory, with non-trivial couplings of the dilaton to matter
\cite{Casas,Polyakov}. Therefore, it is expected that the
description of gravitational phenomena close to Planck scale
should also include this extra scalar field.

The simplest scalar-tensor theory is Jordan--Brans--Dicke (JBD)
theory \cite{JBD} with a constant $\w$ parameter, where the
Brans--Dicke scalar field is in this case the string dilaton,
which acts like a dynamical gravitational constant.
Jordan--Brans--Dicke theory of gravity was introduced in the
inflationary universe scenario as extended inflation \cite{EI},
in an attempt to solve the graceful-exit problem of old
inflation.  However, only small values of the BD parameter were
compatible with observations of the cosmic background radiation,
in contradiction with the large values coming from
post-Newtonian experiments. A possible solution could be a
variable $\w$ parameter \cite{HEI}, see however \cite{ARL}.
Extended Inflation has also been considered in the context of
string effective actions \cite{SEI}. A different proposal is to
consider JBD theory together with chaotic inflation
\cite{ExtChaot}.

In this paper we study the stochastic inflation formalism in the
context of JBD theory of gravity with an arbitrary chaotic
potential for the inflaton. The presence of a scalar field in
the gravitational sector of the theory complicates the picture.
There are new effects associated with the quantum fluctuations
of both the inflaton and dilaton fields; the former defines the
effective dynamics of inflation and the latter the fluctuations
of the gravitational constant \cite{GBLL}.  The main idea of
stochastic inflation is to solve the equations for the
inhomogeneous fields in de Sitter space by separating both the
gravitational and scalar fields in short-distance quantum
fluctuations, which oscillate on scales smaller than the Hubble
radius, and long-distance fluctuations which are treated as
classical fields. It is the special character of de Sitter space
and its horizon which allows this distinction. As the
inflationary domains expand, short wavelengths cross the horizon
and communicate with the long wavelength classical background
fields, acting as a stochastic noise term. Therefore we can
separate the evolution of the long wavelength field as a
two-step process, first a random kick from short-distance
quantum fluctuations (diffusion) followed by classical evolution
(drift). This process can be understood as a Brownian motion of
the fields and thus can be described with the Langevin and
Fokker--Planck equations.

In Section \ref{ADM} we describe the Arnowitt--Deser--Misner
(ADM) \cite{ADM} formalism for JBD theory with an inflaton
scalar field, see also \cite{Luis}. In this Section we follow
closely the work of Salopek and Bond \cite{SB} and generalize it
to JBD. We find the energy and momentum constraints for the
inhomogeneous fields, the gravitational and the scalar fields'
equations. These non-linear equations are extremely difficult to
solve and one usually expands in small perturbations with
respect to a homogeneous background.  Fortunately, in the
context of stochastic inflation, one is allowed to separate the
short-distance behavior from the long-wavelength background
fields, which can then be described by a coarse-grained
homogeneous field. The background field equations are then
obtained by neglecting large-scale gradients.  There is a
natural scale in de Sitter space given by the Hubble radius.
Since we are interested in structures larger than the horizon
scale it is natural to expand in powers of $k/(aH)$ where $k$ is
the comoving wave number of the perturbation. This is done in
Section \ref{SGE}, where we arrive at a consistent set of
background field equations. In Section \ref{inf} we solve these
equations in the synchronous gauge and in the slow-roll
approximation, where we assume that the scalar fields evolve
slower than the expansion of the universe.  This simplifies
significantly the equations and we obtain approximate solutions
for the background fields for generic chaotic potentials of the
type $V(\sig)= {\lam\over2n} \sig^{2n}$. In Section \ref{pert}
we describe the formalism of small perturbations on a
homogeneous background, using linear perturbation theory of the
metric and scalar fields in the longitudinal gauge \cite{MUK}.
We find the amplitude of perturbations in the slow-roll
approximation for the Einstein frame, in terms of both the
dilaton and inflaton fields. Those perturbations have a quantum
mechanical origin in the short-wavelength fluctuations of the
scalar fields. We compute the amplitude of those quantum
fluctuations in a de Sitter universe and find the
Gibbons--Hawking temperature \cite{GibHaw} for both the inflaton
and dilaton in certain well defined limits. We then obtain an
expression for the amplitude of density perturbations in the
cosmic background radiation that should be compared with the one
coming from general relativity.

In Section \ref{SI} we describe the stochastic inflation
formalism \cite{StoInf,SelfRep,GonLin} for Jordan--Brans--Dicke
inflation, see also \cite{GBLL}. The Brownian motion of the
scalar fields in de Sitter space can be written as a Langevin
equation for the coarse-grained fields with an effective white
noise generated by quantum fluctuations. The associated
Fokker--Planck equation for the probability distribution of
finding a given value of the scalar fields in a given point of
space-time can be derived from the Langevin equation and written
in a form which is manifestly time-reparametrization invariant.
The proper way to find stationary solutions is to solve the
diffusion equation subject to certain well defined boundary
conditions \cite{LLM}. We show that the probability distribution
far from the Planck boundary behaves like a Gaussian with fields
centered at their classical trajectories and we calculate its
dispersion coefficients.  The probability distribution in the
physical frame takes into account the exponential growth of each
inflationary domain. The condition for self-reproduction of the
universe is then calculated. Those inflationary
domains with fields inside the region of self-reproduction will
dominate the proper volume of the universe and will tend to
diffuse towards the Planck boundary.

In Section \ref{stat} we study the stationary solutions of the
Fokker--Planck equation in the physical frame by reexpressing
the evolution equation as a Schr\"odinger equation in two
dimensions with an effective potential. We find that the
distribution will very quickly settle at the Planck boundary.
However, since the Planck boundary is a line it will slide along
it, unless we impose extra boundary conditions \cite{GBLL}.  In
Section \ref{conc} we present our conclusions.

\section{\label{ADM} ADM formalism for JBD Cosmology}

In the Arnowitt--Deser--Misner formalism \cite{ADM} the four
dimensional metric $g_{\mu\nu}$ is parame\-trized by the
three-metric $h_{ij}$ and the lapse and shift functions $N$ and
$N^i$, which describe the evolution of time-like hypersurfaces
\cite{MTW,Wald},
\be
g_{00} = - N^2 + h^{ij} N_i N_j, \hs
g_{0i} = g_{i0} = N_i, \hs g_{ij} = h_{ij}.
\eel{GIJ}
The action for the Jordan--Brans--Dicke theory plus an inflaton
scalar field with potential $V(\sig)$ in the ADM formalism has
the form \cite{Luis}
\be\ba{rl}
{\cal S} =&\!{\displaystyle
\int d^4x \sqrt{-g} \left[{\M^2(\phi)\over16\pi} R -
\med(\partial\phi)^2 -\med(\partial\sig)^2 - V(\sig)\right] }
\\[4mm] =&\!{\displaystyle
\int d^4x\ N\sqrt{h} \left[{\M^2(\phi)\over16\pi} \left(^{(3)}\!R +
K_{ij} K^{ij} - K^2 + {4\over\phi} K \Pi^\phi \right)\right.}
\\[4mm] &\!{\displaystyle \left. \hspace{2.3cm}
+\ \med\left[(\Pi^\phi)^2 - \phi_{|i}\phi^{|i}\right]
+ \med\left[(\Pi^\sig)^2 - \sig_{|i}\sig^{|i}\right]
- V(\sig)\right],}
\ea\eel{S}
where ${\displaystyle \M^2(\phi) = {2\pi\over\w} \phi^2 }$ is
the dilaton-dependent Planck mass and $\Pi^{\phi_k}$ are the
scalar-fields' momenta
\be\ba{c}
{\displaystyle
\Pi^\phi = {1\over N}(\dot\phi - N^i \phi_{|i}) , }\\[4mm]
{\displaystyle
\Pi^\sig = {1\over N}(\dot\sig - N^i \sig_{|i}) . }
\ea\eel{SFM}
Vertical bars denote three-space-covariant derivatives with
connections derived from $h_{ij}$, $^{(3)}\!R$ is the
three-space curvature associated with the metric $h_{ij}$, and
$K_{ij}$ is the extrinsic curvature three-tensor
\be
K_{ij} = {1\over2N} (N_{i|j} + N_{j|i} - \dot h_{ij}),
\eel{KIJ}
where a dot denotes differentiation with respect to the time
coordinate. The traceless part of a tensor is denoted by an
overbar.  In particular,
\be
\bar K_{ij} = K_{ij} - {1\over3} K h_{ij}, \hs K = K_i^i.
\eel{TRC}
The trace $K$ is a generalization of the Hubble parameter, as
will be shown below. Variation of the action with respect to $N$
and $N^i$ yields the energy and momentum constraint equations
\be\ba{c}
{\displaystyle
- ^{(3)}\!R + \bar K_{ij} \bar K^{ij} - {2\over3} K^2 +
{4\over\phi} K \Pi^\phi + {16\pi\over\M^2(\phi)} \eps
= 0 ,}\\[4mm]
{\displaystyle
\bar K^j_{i|j} - {2\over3} K_{|i} + {2\over\phi}\left(
\phi_{|j} \bar K^j_i + {1\over3} \phi_{|i} K\right)
+ {2\over\phi^2} (\phi \Pi^\phi)_{|i}
+ {8\pi\over\M^2(\phi)} \left(\Pi^\phi \phi_{|i} +
\Pi^\sig \sig_{|i}\right) = 0 .}
\ea\eel{EMC}

Variation with respect to $h_{ij}$ gives the dynamical
gravitational field equations, which can be separated into the
trace and traceless parts
\be\ba{rl}
\dot K - N^i K_{|i} =&\!{\displaystyle
-\ N^{|i}_{|i} + N \left({1\over4}c^{(3)}\!R + {3\over4} \bar
K_{ij} \bar K^{ij} + \med K^2 + {4\pi\over\M^2(\phi)} T\right)
}\\[4mm]
&\!{\displaystyle
-\ {2\over\phi} K N \Pi^\phi + {3\over\phi} (\dot\Pi^\phi
- N^i \Pi^\phi_{|i}) + {3N\over\phi^2} (\Pi^\phi)^2 }
\ea\eel{TRA}
\be\ba{rl}
\dot{\bar K^i}_j - N^k \bar K^i_{j|k} &\!{\displaystyle
+\ N^i_{|k} \bar K^k_j - N^k_{|j} \bar K^i_k = \ - N^{|i}_{|j}
+ {1\over3} N^{|k}_{|k} \delta^i_j  }\\[4mm]
&\!{\displaystyle
-\ {2\over\phi} \bar K^i_j N \Pi^\phi +
N \left(^{(3)}\!\bar R^i_j + K \bar K^i_j
- {8\pi\over\M^2(\phi)} \bar T^i_j \right) .}
\ea\eel{TRL}

Variation with respect to $\phi$ and $\sig$ gives the
scalar-fields' equations of motion
\be\ba{rl}
{\displaystyle
{1\over N}(\dot\Pi^\phi - N^i \Pi^\phi_{|i}) }
&\!{\displaystyle  -\ K \Pi^\phi
- {1\over N} N_{|i} \phi^{|i} - \phi_{|i}^{\ |i}
+ {\M^2(\phi)\over4\pi\phi N} (\dot K - N^i K_{|i})}\\[3mm]
&\!{\displaystyle
-\ {\M^2(\phi)\over8\pi\phi} \left(^{(3)}\!R + \bar K_{ij}
\bar K^{ij} + {4\over3} K^2 \right) = 0,}\\[4mm]
{\displaystyle
{1\over N}(\dot\Pi^\sig - N^i \Pi^\sig_{|i}) }
&\!{\displaystyle  -\ K \Pi^\sig
- {1\over N} N_{|i} \sig^{|i} - \sig_{|i}^{\ |i} +
{\partial V\over\partial\sig} = 0 }.
\ea\eel{SFE}
The energy density on a constant-time hypersurface is
\be
\eps = \med \left[(\Pi^\phi)^2 + \phi_{|i} \phi^{|i}\right] +
\med \left[(\Pi^\sig)^2 + \sig_{|i} \sig^{|i}\right] + V(\sig),
\eel{EDY}
and the stress three-tensor
\be
T_{ij} = \phi_{|i} \phi_{|j} + \sig_{|i} \sig_{|j} + h_{ij}
\left[\med \left[(\Pi^\phi)^2 - \phi_{|k} \phi^{|k}\right] +
\med \left[(\Pi^\sig)^2 - \sig_{|k} \sig^{|k}\right]
- V(\sig)\right].
\eel{TIJ}

It is extremely difficult to solve these highly nonlinear
coupled equations in a cosmological scenario without making some
approximations. The usual approach is to assume homogeneity of
the fields to give a background solution and then linearize the
equations to study deviations from spatial uniformity. The
smallness of cosmic microwave background anisotropies
\cite{COBE} gives some justification for this perturbative
approach at least in our local part of the Universe. However,
there is no reason to believe it will be valid on much larger
scales. In fact, the stochastic approach to inflation suggests
that the Universe is extremely inhomogeneous on very large
scales \cite{LinBook}.  Fortunately, in this framework one can
coarse-grain over a horizon distance and separate the short-
from the long-distance behavior of the fields, where the former
communicates with the latter through stochastic forces. The
equations for the long-wavelength background fields are obtained
by neglecting large-scale gradients, leading to a
self-consistent set of equations, as we will discuss in the
next section.

\section{\label{SGE} Spatial gradient expansion}

It is reasonable to expand in spatial gradients whenever the
forces arising from time variations of the fields are much
larger than forces from spatial gradients. In linear
perturbation theory one solves the perturbation equations for
evolution outside of the horizon: a typical time scale is the
Hubble time $H^{-1}$, which is assumed to exceed the gradient
scale $a/k$, where $k$ is the comoving wave number of the
perturbation.
Since we are interested in structures on scales larger than the
horizon, it is reasonable to expand in $k/(aH)$. In particular,
for inflation this is an appropriate parameter of expansion
since spatial gradients become exponentially negligible after a
few $e$-folds of expansion beyond horizon crossing, $k=aH$.

It is therefore useful to split the fields $\phi$ and $\sig$
into coarse-grained long-wavelength background fields
$\phi(t,x^j)$ and residual short-wavelength fluctuating fields
$\delta\phi(t,x^j)$.  There is a preferred timelike hypersurface
within the stochastic inflation approach in which the splitting
can be made consistently, but the definition of the background
field will depend on the choice of hypersurface, {\em i.e.} the
smoothing is not gauge invariant. For stochastic inflation the
natural smoothing scale is the comoving Hubble length
$(aH)^{-1}$ and the natural hypersurfaces are those on which
$aH$ is constant.  In that case a fundamental difference between
between $\phi$ and $\delta\phi$ is that the short-wavelength
components are essentially uncorrelated at different times,
while long-wavelength components are deterministically
correlated through the equations of motion.

In order to solve the equations for the background fields, we
will have to make suitable approximations. The idea is to expand
in the spatial gradients of $\phi$ and to treat the terms that
depend on the fluctuating fields as stochastic forces describing
the connection between short- and long-wavelength components.
In this Section we will neglect the stochastic forces due to
quantum fluctuations of the scalar fields and will derive the
approximate equation of motion for the background fields.  We
retain only those terms that are at most first order in spatial
gradients, neglecting such terms as $\phi_{|i}^{\ |i}$,
$\phi_{|i} \phi^{|i}$, $^{(3)}\!R$, $^{(3)}\!R_i^j$, and
$\bar T_i^j$.

We will also choose the simplifying gauge $N^i=0$. The evolution
equation \form{TRL} for the traceless part of the extrinsic
curvature is then $\dot{\bar K^i}_j = N K \bar K^i_j -
{2\over\phi} \bar K^i_j \dot\phi$. Using ${\displaystyle N K = -
{\partial\over \partial t} \ln\sqrt h}$ from \form{TRC}, we find
the solution $\bar K^i_j \propto \phi^{-2}\ h^{-1/2}$, where $h$
is the determinant of $h_{ij}$. During inflation $h^{-1/2}\equiv
a^{-3}$, with $a$ the overall expansion factor, therefore $\bar
K^i_j$ decays extremely rapidly and can be set to zero in the
approximate equations. The most general form of the three-metric
with vanishing $\bar K^i_j$ is
\be
h_{ij} = a^2(t,x^k)\ \gamma_{ij}(x^k), \hs a(t,x^k)\equiv
\exp[\alpha(t,x^k)] ,
\eel{HIJ}
where the time-dependent conformal factor is interpreted as a
space-dependent expansion factor. The time-independent
three-metric $\gamma_{ij}$, of unit determinant, describes the
three-geometry of the conformally transformed space.  Since
$a(t,x^k)$ is interpreted as a scale factor, we can substitute
the trace $K$ of the extrinsic curvature for the Hubble
parameter
\be
H(t,x^i) \equiv {1\over N(t,x^i)}\ \dot\alpha(t,x^i) =
-{1\over3} K(t,x^i).
\eel{HUB}

Furthermore, since the Brans--Dicke parameter $\w$ is bounded by
post-Newtonian experiments \cite{VIK} and primordial
nucleosynthesis \cite{PNB} to be very large, $\w > 500$, we will
use the approximation $\w \gg 1$ in equations
(\ref{EMC}--\ref{SFE}).  The energy and momentum constraint
equations \form{EMC} can now be written as
\be\ba{c}
{\displaystyle
H^2 = {8\pi\over3\M^2(\phi)} \left(\med (\Pi^\phi)^2 + \med
(\Pi^\sig)^2 + V(\sig)\right) - {2\over\phi} H\Pi^\phi,}\\[4mm]
{\displaystyle
H_{|i} = - {4\pi\over\M^2(\phi)} (\Pi^\phi \phi_{|i} +
\Pi^\sig \sig_{|i}) + {H\over\phi} \phi_{|i} ,}
\ea\eel{MEC}
together with the evolution equation \form{TRA}
\be
- {1\over N} \dot H = {3\over2} H^2 + {4\pi\over3\M^2(\phi)} T
- {1\over\phi} H \Pi^\phi ,
\eel{ART}
where $T = 3 \left(\med (\Pi^\phi)^2 + \med (\Pi^\sig)^2 -
V(\sig)\right)$.

In general, $H$ is a function of the scalar fields and time,
$H(t,x^i) \equiv H(\phi(t,x^i),\sig(t,x^i),t)$. From the momentum
constraint \form{MEC} we find that the scalar-fields' momenta must
obey
\be\ba{rl}
\Pi^\phi =&\!{\displaystyle
-\ {\M^2(\phi)\over4\pi} \left[\left({\partial H\over\partial\phi}
\right)_t - {H\over\phi}\right] ,}\\[4mm]
\Pi^\sig =&\!{\displaystyle
-\ {\M^2(\phi)\over4\pi} \left({\partial H\over
\partial \sig}\right)_t .}
\ea\eel{PI}
Contrary to the situation in Einstein theory \cite{SB}, $H$ has
an explicit time dependence. Comparing equation \form{ART} with
the time derivative of
$H(\phi,\sig,t)$,
\be\ba{rl}
{\displaystyle
{1\over N} \left({\partial H\over\partial t}\right)_x }
=&\!{\displaystyle
\Pi^\phi \left({\partial H\over\partial\phi}\right)_t +
\Pi^\sig \left({\partial H\over\partial\sig}\right)_t +
{1\over N} \left({\partial H\over\partial t}\right)_{\phi,\sig}
}\\[4mm]
=&\!{\displaystyle
{H\over\phi} \Pi^\phi - {4\pi\over\M^2(\phi)}
\left((\Pi^\phi)^2 + (\Pi^\sig)^2\right) + {1\over N}
\left({\partial H\over\partial t}\right)_{\phi,\sig} ,}
\ea\eel{HXT}
we find ${\displaystyle {1\over N} \left({\partial H\over
\partial t}\right)_{\phi,\sig} = {3\over\phi} H \Pi^\phi}$.
It is only in the slow-roll approximation, when we neglect
$\Pi^\phi$ versus $H$, that the Hubble parameter becomes
time-independent.

On the other hand, the scalar fields' equations \form{SFE}
can be written to first order in spatial gradients as
\be\ba{rl}
{\displaystyle
{1\over N}\dot\Pi^\phi + 3 H \Pi^\phi\ =}&\!{\displaystyle
{3\M^2(\phi)\over2\pi} H^2 ,}\\[4mm]
{\displaystyle
{1\over N}\dot\Pi^\sig + 3 H \Pi^\sig\ =}&\!{\displaystyle
-\ {\partial V\over\partial\sig} }.
\ea\eel{FES}
This set of equations are still too complicated to solve for
arbitrary potentials $V(\sigma)$. In the next section we will
find solutions to them in the slow-roll approximation.

\section{\label{inf} Jordan--Brans--Dicke Inflationary Cosmology}

We are now interested in the classical behavior of the
long-wavelength quasi-homogeneous background fields, and later
will study the effect of the stochastic perturbations on these
background fields. Furthermore, we will chose the synchronous
gauge $N=1$, together with $N^i=0$, and solve the homogeneous
part of the inflationary equations of motion \form{FES},
\form{MEC} in the slow-roll approximation,
\be
H^2 = H^2_{\rm SR} \equiv {8\pi V(\sig)\over3\M(\phi)} ,
\eel{HSR}
together with $\ddot\phi \ll H\dot\phi \ll H^2 \phi$. The
field equations then look like \cite{ExtChaot}
\be\ba{rl}
&{\displaystyle \dot\phi =
-\ {\M^2(\phi)\over2\pi} {\partial H\over\partial\phi} =
\ \frac{\phi}{\w} H , }\\[4mm]
&{\displaystyle \dot\sig =
-\ {\M^2(\phi)\over4\pi} {\partial H\over\partial\sig} =
- \frac{V'(\sig)}{3H} , }\\[4mm]
&\dot\alpha = H
\ea\eel{SEQ}
Let us now study the behavior of the approximate inflationary
solutions for a generic chaotic potential, $V(\sig) = {\lam
\over2n} \sig^{2n}$,
\be\ba{rl}
&{\displaystyle
\dot{\phi} = \frac{2}{n} \left(\frac{n\lam}
{6\w}\right)^{1/2} \sig^n },\\[3mm]
&{\displaystyle
\dot{\sig} = -\frac{\phi}{\sig} \left(
\frac{n\lam}{6\w}\right)^{1/2} \sig^n }.
\ea\eel{EVO}
For these theories, $\phi$ and $\ \varphi\equiv \sqrt{2\over n}
\ \sig $ move along a circumference of constant radius in the
plane $(\varphi,\phi)$. We can parametrize the classical
trajectory by polar coordinates $(\phi(t), \varphi(t)) =
(r\sin\theta(t), r\cos\theta(t))$, with constant radius $r$,
and angular velocity given by
\be
\dot{\theta}(t) = \left(\frac{\lam}{3\w}\right)^{1/2}
\left(\frac{n}{2}\right)^{\frac{n-1}{2}}
[r\cos\theta(t)]^{n-1}.
\eel{DXT}
For $n=1$ we find a constant angular velocity and the solutions
can be written as \cite{ExtChaot}
\be\ba{rl}
&{\displaystyle  \phi(t) =
r \sin\left(\theta_o + \frac{m t}{\sqrt{3\w}}
\right) },\\[3mm]
&{\displaystyle  \sig(t) =
\frac{r}{\sqrt{2}} \cos\left(\theta_o + \frac{m t}
{\sqrt{3\w}}\right) },\\[4mm]
&{\displaystyle  a(t) =
a_o \left(\frac{\sin\left(\theta_o + \frac{m t}{\sqrt{3\w}}
\right)}{\sin\theta_o}\right)^\w \sim t^\w, \hs \theta_o
\frac{\sqrt{3\w}}{m} < t \ll {\pi\sqrt{3\w}\over2m} },
\ea\eel{CLA}
with the usual power-law behavior of extended inflation. For
$n\geq2$, the angular velocity decreases and the classical
solutions are more complicated, although still power-law. These
solutions are actually attractors of the complete equations of
motion, for all $n$ \cite{DGL}.

In the chaotic inflation scenario, the most natural initial
conditions for inflation are set at the Planck boundary,
$V(\sig_p) \simeq \M^4(\phi_p)$, beyond which a classical
space--time has no meaning and the energy gradient of the
inhomogeneities produced during inflation becomes greater than
the potential energy density, thus preventing inflation itself.
The initial conditions for inflation are thus defined at the
curve $\phi_p^2 = {\w\over2\pi} \sqrt{\lam\over2n} \sig_p^n$.
On the other hand, inflation will end when the kinetic energy
density of the scalar fields becomes comparable with the
potential energy density, $\med\dot{\phi}^2 + \med\dot{\sig}^2
\simeq V(\sig)$ or $\sig_e = {n\over2\sqrt{3\pi}} \M(\phi_e) =
{n\over\sqrt{6\w}} \phi_e$.

In the absence of any potential for $\phi$, the dilaton remains
approximately constant after inflation, and therefore the Planck
mass today is given by its value at the end of inflation,
${\displaystyle \M \simeq \sqrt{2\pi\over\w}\ \phi_e }$, while
the total amount of inflation is approximately ${\displaystyle
{a_e\over a_p}\simeq \left({\phi_e\over\phi_p} \right)^\w }$.

Now that we have the background solutions we would like to
describe the formalism of small perturbations on the homogeneous
background, to account for the quantum fluctuations in de Sitter
space.

\section{\label{pert} Perturbations on a homogeneous background}

In order to compute the perturbations on a homogeneous
long-wavelength background, we use linear perturbation theory of
the metric and scalar fields in the longitudinal gauge and apply
it to our case of Brans-Dicke theory with an inflaton field. We
follow the notation of \cite{MUK}, and define
\be\ba{rl}
\phi(t,x^i) =&\!\phi(t) + \delta\phi(t,x^i) \\[3mm]
\sig(t,x^i) =&\!\sig(t) + \delta\sig(t,x^i) \\[3mm]
N(t,x^i) =&\!1 + \Phi(t,x^i) ,\hs \ N^i(t,x^i) = 0\\[3mm]
h_{ij}(t,x^i) =&\![1-2\Phi(t,x^i)]\ a^2(t)\ \delta_{ij} \ .
\ea\eel{NOT}
Note that we are calling $\phi$ to the homogeneous part, to
avoid carrying unnecessary subindices, and $\delta\phi$ to the
inhomogeneous perturbation. The extrinsic curvature \form{KIJ}
and the three-curvature scalar take simple expressions,
\be
K = - 3 H (1 - \Phi) + 3 \dot\Phi, \hs
\bar K_{ij} = 0 , \hs \ ^{(3)}\!R = {4\over a^2}
\Phi_{|i}^{\ |i} \ ,
\eel{K3R}
while the scalar fields' momenta are $\Pi^\phi = \dot\phi +
\dot{\delta\phi} - \dot\phi\Phi$. The homogeneous part of the
momentum constraint \form{EMC} is trivial, but that of the
energy constraint is not,
\be
H^2 = {8\pi\over3\M^2(\phi)} \left(\med \dot\phi^2 + \med \dot\sig^2
+ V(\sig)\right) - 2 H {\dot\phi\over\phi}\ .
\eel{HEC}
The inhomogeneous parts of the energy and momentum constraints
have complicated expressions,
\be\ba{c}
{\displaystyle
{1\over a^2} \Phi_{|i}^{\ |i} - 3 H^2 \Phi
- 3 H \dot\Phi = 3 H
{\dot\phi\over\phi} \left(2\Phi + {\delta\phi\over\phi} -
{\dot{\delta\phi}\over\dot\phi}\right) + 3 {\dot\phi\over\phi}
\dot\Phi }\\[4mm]
{\displaystyle + {4\pi\over\M^2(\phi)} \left(\dot\phi\dot{\delta\phi}
+ \dot\sig\dot{\delta\sig} - \dot\phi^2 \Phi - \dot\sig^2 \Phi
- \dot\phi^2 {\delta\phi\over\phi} - \dot\sig^2 {\delta\phi\over\phi}
+ {\partial V\over\partial\sig} \delta\sig - 2 V
{\delta\phi\over\phi}
\right)\ , }
\ea\eel{IEC}
\be
\left(\dot\Phi + H \Phi +
H {\delta\phi\over\phi}\right)_{|i} =
{4\pi\over\M^2(\phi)} \left(\dot\phi\delta\phi +
\dot\sig\delta\sig\right)_{|i} \ .
\eel{IMC}

The homogeneous part of the gravitational equation \form{TRA} is
\be
2 \dot H + 3 H^2 = - {8\pi\over\M^2(\phi)} \left( \med
\dot\phi^2 + \med \dot\sig^2 - V(\sig)\right) + 2 H
{\dot\phi\over\phi}.
\eel{HGE}
while the inhomogeneous part can be written as
\be\ba{c}
{\displaystyle
 \ddot\Phi + 4 H \dot\Phi + 2 \dot H \Phi + 3 H^2 \Phi = H
{\dot\phi\over\phi} \left(2\Phi + {\delta\phi\over\phi} -
{\dot{\delta\phi}\over\dot\phi}\right) + {\dot\phi\over\phi}
\dot\Phi + {1\over a^2\phi} \delta\phi_{|i}^{\ |i} }\\[4mm]
{\displaystyle + {4\pi\over\M^2(\phi)} \left(\dot\phi\dot{\delta\phi}
+ \dot\sig\dot{\delta\sig} - \dot\phi^2 \Phi - \dot\sig^2 \Phi -
\dot\phi^2 {\delta\phi\over\phi} - \dot\sig^2 {\delta\phi\over\phi}
- {\partial V\over\partial\sig} \delta\sig
+ 2 V {\delta\phi\over\phi}\right), }
\ea\eel{IGE}
Putting together \form{HEC} and \form{HGE}, we find
\be
\dot H = {\ddot a\over a} - H^2 = - {4\pi\over\M^2(\phi)}
\left(\dot\phi^2 + \dot\sig^2\right) + 4 H {\dot\phi\over\phi}.
\eel{DDA}

Furthermore, the scalar field equations can be separated into
homogeneous and inhomogeneous parts
\be\ba{rl}
{\displaystyle \ddot\phi + 3 H \dot\phi }=&\!
{\displaystyle {3\phi\over\w} H^2 }\\[4mm]
{\displaystyle \ddot\sig + 3 H \dot\sig }=&\!
{\displaystyle -\ {\partial V\over\partial\sig}  }.
\ea\eel{HSF}
\be\ba{rl}
{\displaystyle \ddot{\delta\phi} + 3 H \dot{\delta\phi}
- {1\over a^2} \delta\phi_{|i}^{\ |i} - 4 \dot\phi \dot\Phi}=&\!
{\displaystyle {3\over\w} \left({\phi\over12a^2} \Phi_{|i}^{\ |i} +
H^2 \delta\phi - 2 \phi H\dot\Phi \right) }\\[4mm]
{\displaystyle \ddot{\delta\sig} + 3 H \dot{\delta\sig}
- {1\over a^2} \delta\sig_{|i}^{\ |i} - 4 \dot\sig \dot\Phi}=&\!
{\displaystyle -\ {\partial^2 V\over\partial\sig^2} \delta\sig
- 2 {\partial V\over\partial\sig} \Phi }.
\ea\eel{ISF}

Subtracting \form{IGE} from \form{IEC} and using equations
\form{IMC}, \form{DDA}, and \form{HSF}, one finally
finds a long wavelength solution very similar to that of
reference \cite{MUK},
\be
\Phi + {\delta\phi\over\phi} = A \left(1 - {\dot a\over a^2}
\int a dt\right) ,
\eel{PAT}
where $A$ is a constant that can be calculated during inflation,
using \form{IMC} and \form{DDA} in the slow-roll approximation, as
\be
A = {a\over\int a dt} \left({\dot\phi\delta\phi +
\dot\sig\delta\sig\over\dot\phi^2 + \dot\sig^2}\right).
\eel{A}

In the Einstein frame $\bar g_{\mu\nu} = \phi^2 g_{\mu\nu}$, the
density contrast of perturbations that enter the horizon during
the matter era is related to both $\Phi$ and $\delta\phi$ by
\cite{MUK} \footnote{This expression was first obtained in
\cite{EDP} in the Einstein frame. See also \cite{DGL}.}
\be
{\delta\bar\rho\over\bar\rho} = - 2 \left(\Phi + {\delta\phi
\over\phi}\right) = - {6\over5} H {\dot\phi\delta\phi +
\dot\sig\delta\sig\over\dot\phi^2 + \dot\sig^2} .
\eel{DRR}

The energy density perturbations we observe in the cosmic
background radiation could have originated during inflation in
our model \form{S}, as quantum fluctuations of both scalar
fields that first left the horizon during inflation and later
reentered during the radiation or matter dominated eras. The
amplitude of energy density perturbations \form{DRR} associated
with a given wavelength should be evaluated during inflation at
the time in which that wavelength first crossed the horizon.

\subsection{Quantum Fluctuations of Scalar Fields}

We will now compute the amplitude $\delta\phi$ and
$\delta\sigma$ of the quantum fluctuations of the scalar fields
in de Sitter space, whose wavelengths are stretched beyond the
horizon and act on the quasi--homogeneous background fields like
an stochastic force. In order to calculate this effect we
coarse--grain over a horizon distance and separate the scalar
fields into long--wavelength classical background fields
$\phi(x)$ and $\sig(x)$ plus short--wavelength quantum
fluctuations $\delta\phi(x)$ and $\delta\sig(x)$ with physical
momenta $k/a > H$,
\be\ba{rl}
&{\displaystyle
\delta\phi(\vx,t) = \int d^3k \ \theta(k -
\varepsilon a H)\left[a_k u_k(x) + a^\dagger_k u^\ast_k(x)
\right] },\\[3mm]
&{\displaystyle
\delta\sig(\vx,t) = \int d^3k' \ \theta(k'-
\varepsilon a H)\left[b_{k'}v_{k'}(x)+ b^\dagger_{k'}
v^\ast_{k'}(x)\right] },
\ea\eel{CGF}
where $\varepsilon$ is an arbitrary parameter that shifts the
scale for coarse--graining \cite{StoInf}. Physical results turn
out to be independent of the choice of $\varepsilon$. The
quantum fluctuations are assumed to satisfy the following
commutation relations
\begin{equation}\label{COM}
[a_k, a^\dagger_{k'}] = [b_k, b^\dagger_{k'}] =
\delta^3(\vec k - \vec k') \ , \hspace{6mm}
[a_k, b^\dagger_{k'}] = 0 \ .
\end{equation}
The approximate solutions to the perturbation equations
\form{ISF} in de Sitter space with $V(\sig) = \med m^2 \sig^2$
are given by \cite{BD}
\be\ba{rl}
u_k(x)=&\!{\displaystyle
{e^{i\vec{k}\cdot\vec{x}}\over(2\pi)^{3/2}}\ {H\eta\over2}
\sqrt{\pi\eta}\ H_\mu^{(2)}(k \eta)\ , \hspace{1.2cm}
\mu^2 = \frac{9}{4} + \frac{3}{\omega} \simeq \frac{9}{4} }\ ,\\[4mm]
v_{k'}(x)=&\!{\displaystyle
{e^{i\vec{k'}\cdot\vec{x}}\over(2\pi)^{3/2}}\ {H\eta\over2}
\sqrt{\pi\eta}\ H_\nu^{(2)}(k'\eta)\ , \hspace{1cm}
\nu^2 = \frac{9}{4} - \frac{m^2}{H^2} \simeq \frac{9}{4} }\ ,
\ea\eel{UVK}
where $\eta = - (a H)^{-1}$ is the conformal time, and
$H^{(2)}_{3/2}(x) = - \sqrt{2\over\pi x}\left(1+ {i\over
x}\right) \exp(ix)$.

The amplitude of the quantum fluctuations of $\phi$ and
$\sigma$ can then be computed as
\be\ba{c}
{\displaystyle
\delta\phi = \left(4\pi k^3 |u_k|^2\right)^{1/2} =
{H\over2\pi}, }\\[3mm]
{\displaystyle
\delta\sig = \left(4\pi k'^3 |v_{k'}|^2\right)^{1/2} =
{H\over2\pi}, }
\ea\eel{STEP}
which coincides with the Gibbons--Hawking temperature
\cite{GibHaw}. It should be noted that we have made two
approximations when deriving this formula, first that $\mu$ and
$\nu$ are both $3/2$, see \form{UVK}, and second that $H$ is
constant, when in general this is not true. We should have
corrections to \form{STEP}, but they are expected to be small
\cite{SB}.

Substituting the amplitude of quantum fluctuations of the scalar
fields \form{STEP} into the energy density perturbations
\form{DRR} and using the equations of motion \form{SEQ}, we find
for arbitrary $V(\sigma)$
\be
\left.{\delta\bar\rho\over\bar\rho} = {24\over5}
{H(\sig,\phi)\over\M^2(\phi)}{V(\sig)\over V'(\sig)}
\left({1+\dot\phi/\dot\sigma\over1+(\dot\phi/\dot\sig)^2}
\right)\right|_{N_\lam},
\eel{PER}
where $N_\lambda$ stands for the number of e-folds before the
end of inflation associated with the horizon crossing of a
particular wavelength.  For perturbations of the size of the
present horizon, we must compute the last expression at
$N_\lambda \sim 65$ \cite{LinBook}.  Note that in the large $\w$
limit, $\dot{\phi}\ll\dot{\sig}$ during the last stages of
inflation, which ensures the approximate equivalence of the
Einstein and Jordan frame. We then recover the usual expression
\cite{LinBook}
\be
\left.{\delta\rho\over\rho} \simeq {24\over5}
{H(\sig,\phi)\over\M^2(\phi)}{V(\sig)\over V'(\sig)}
\right|_{N_\lam},
\eel{PRE}
where $\M$ is now $\phi$--dependent. For theories with potentials
of the type ${\lambda\over2n} \sigma^{2n}$, it behaves like
\be
\left.{\delta\rho\over\rho} \simeq
{6\w\over5n\pi} \left({2\w\lam\over3n}\right)^{1/2}
{\sig^{n+1}\over\phi^3}\right|_{N_\lam}.
\eel{REP}
In the case of the theory $\lam\sig^4/4$, the density perturbation
\form{REP} takes the usual $\sqrt{\lam}$ dependence. However,
in the case $m^2\sig^2/2$, the perturbation can be written as
\be
{\delta\rho\over\rho} \simeq {2e^{3N_\lam\over\w}\over5\sqrt{3\pi}}
\left[1+3\w\left(1-e^{-{2N_\lambda\over\w}}
\right)\right] {m\over\M(\phi_e)}
\eel{RPE}
Therefore, we note that the larger is the Planck mass at the end
of inflation in a given region of the Universe, the smaller is
the density perturbation in this region. For a detailed discussion
of its cosmological implications see ref. \cite{GBLL}.

\section{\label{SI} Stochastic Inflation}

We can now put together the results of previous sections and
study the Brownian motion of the background fields under the
stochastic impulses of their own quantum fluctuations. We could
write the evolution of the coarse--grained fields in the form of
Langevin equations
\be\ba{rl}
&\Delta\phi = \Pi^\phi N \Delta t + \Delta S_\phi\ ,\\[3mm]
&\Delta\sig = \Pi^\sig N \Delta t + \Delta S_\sig\ ,\\[3mm]
&\Delta\alpha = H N \Delta t + \Delta S_\alpha\ ,
\ea\eel{LAN}
where $\Delta S_{\phi_k}$ are the stochastic noise terms acting
on the long-wavelength background. They are Gaussian distributed
with zero mean and variance
\be
\left\langle\Delta S_{\phi_i}(t)\ \Delta S_{\phi_j}(t')
\right\rangle = \delta_{ij} \left({H(\phi_k)\over2\pi}
\right)^2 H(\phi_k) N \Delta t\ \delta(t-t'),
\eel{VAR}
which corresponds to the Brownian motion of the background
scalar fields $\phi$ and $\sigma$ with step \form{STEP}.  It is
then straightforward to derive the corresponding Fokker--Planck
equation, see e.g. the appendix in \cite{SB},
\be\ba{rl}
{\displaystyle \frac{\partial P_c}{\partial t}}=
&\!{\displaystyle \frac{\partial}{\partial\sig}
\left(\frac{\M^2}{4\pi}\frac{\partial H}{\partial\sig} N P_c
+{(N H^3)^{1/2}\over8\pi^2}
{\partial\over\partial\sig}\left((N H^3)^{1/2}
P_c\right)\right) }\\[3mm]
+&\!{\displaystyle \frac{\partial}{\partial\phi}
\left(\frac{\M^2}{2\pi}\frac{\partial H}{\partial\phi} N P_c
+{(N H^3)^{1/2}\over8\pi^2} {\partial\over\partial\phi}
\left((N H^3)^{1/2}P_c\right)\right)
\equiv - \frac{\partial J_\phi}{\partial\phi} -
\frac{\partial J_\sig}{\partial\sig} },
\ea\eel{FPE}
where we have chosen the Stratonovich version of stochastic
processes. $P_c(\sig,\phi;t)$ represents the probability
distribution of finding a given field configuration
$(\sig,\phi)$ at a given point (of size $H^{-1}$) in comoving
space.  Equation \form{FPE} can be interpreted as the continuity
equation ${\displaystyle \frac{\partial P}{\partial t} +
\nabla\cdot J = 0 }$, associated with the conservation of
probability.  The first terms of each current correspond to the
classical drift forces \form{SEQ} associated with the scalar
fields, while the second terms correspond to the quantum
diffusion due to short--wavelength fluctuations \form{STEP}.

It can be seen by inspection that there is an exact time
reparametrization invariant solution given by
\be
P_c(\sigma,\phi;t) \sim (N H^3)^{-1/2}\ \exp
\left({3\M^4(\phi)\over8V(\sigma)}\right) \ ,
\eel{HH2}
which is proportional to the square of the Hartle--Hawking wave
function of the Universe \cite{HH}, for an arbitrary potential
$V(\sigma)$ and a variable gravitational constant. Unfortunately,
this function is not normalizable for potentials that vanish at
their minimum. But the real problem is that the inflationary
regime ends much before the fields reach the minimum of the
potential, and thus equation \form{FPE} is not valid there.
Therefore, there are no stationary solutions for $P_c$ in
realistic inflationary models. The proper way to find stationary
solutions during the inflationary phase would be to consider
the distribution $P_p$ and solve the corresponding diffusion
equation subject to certain boundary conditions \cite{LLM}.
This was done explicitly in ref. \cite{GBLL}; we will return
to this question in the next section.

Instead of looking for stationary solutions of equation \form{FPE} we
will try to understand the general behavior of the probability
distribution $P_c(\sig,\phi;t)$ in the synchronous gauge $N=1$,
for generic potentials $V(\sig)={\lam\over2n}\sig^{2n}$.  $P_c$
then satisfies the equation
\be\ba{rl}
{\displaystyle \frac{\partial P_c}{\partial t}}=
&\!{\displaystyle
\frac{\partial}{\partial\sig} \left({1\over2\w} \left(
{2\w\lam\over3n}\right)^{1/2} \phi^2
{\partial h\over\partial\sig} P_c + {1\over8\pi^2} \left(
{2\w\lam\over3n}\right)^{3/2} h^{3/2}
{\partial\over\partial\sig}\left(h^{3/2}P_c\right)\right)
}\\[3mm]
+&\!{\displaystyle
\frac{\partial}{\partial\phi} \left({1\over\w} \left(
{2\w\lam\over3n}\right)^{1/2} \phi^2
{\partial h\over\partial\phi} P_c + {1\over8\pi^2} \left(
{2\w\lam\over3n}\right)^{3/2} h^{3/2}
{\partial\over\partial\phi}\left(h^{3/2}P_c\right)\right) }
\ea\eel{FPC}

In order to follow the evolution of the scalar fields under
classical drift forces and quantum diffusion processes, it
will be useful to introduce some redefinitions,
\be
h = {\sig^n\over\phi}, \hs
u=\frac{1}{8\pi^2}\left(\frac{2\w\lam}{3n}\right)^{3/2}t.
\eel{XYH}
We will follow closely the work of Goncharov and Linde
\cite{GonLin}, generalized to JBD. We define the new functions
$A$ and $B$ as
\be
A_\sig = - \frac{3n}{2\lam} \left({2\pi\over\w}\right)^2
\phi^2 \frac{\partial h}{\partial \sig}(\sig,\phi), \hs
A_\phi = - \frac{3n}{\lam} \left({2\pi\over\w}\right)^2
\phi^2 \frac{\partial h}{\partial \phi}(\sig,\phi), \hs
B = h^3(\sig,\phi),
\eel{AAB}
and write the evolution equation for the probability
distribution,
\be\ba{rl}
{\displaystyle
{\partial P_c\over\partial u}}=
&\!{\displaystyle \frac{\partial}{\partial \sig}
\left[-A_\sig(\sig,\phi) P_c(\sig,\phi) +
B^{1/2}(\sig,\phi) \frac{\partial}{\partial\sig} \left(
B^{1/2}(\sig,\phi)P_c(\sig,\phi) \right)\right]}\\[3mm]
+&\!{\displaystyle \frac{\partial}{\partial\phi}
\left[-A_\phi(\sig,\phi) P_c(\sig,\phi) +
B^{1/2}(\sig,\phi) \frac{\partial}{\partial\phi} \left(
B^{1/2}(\sig,\phi)P_c(\sig,\phi) \right)\right] }.
\ea\eel{NEV}

We will now consider the case in which $V(\sig)\ll\M^4(\phi)$,
where quantum diffusion is much smaller than classical drift,
\be
\frac{\partial}{\partial\sig} B(\sig,\phi) \ll |A_\sig(\sig,\phi)|,
\hs
\frac{\partial}{\partial\phi} B(\sig,\phi) \ll |A_\phi(\sig,\phi)|,
\eel{DRF}
and the classical solutions to the equations of motion satisfy
$\dot{\sig}_c(u) = A_\sig(\sig_c,\phi_c)$ and $\dot{\phi}_c(u) =
A_\phi(\sig_c,\phi_c)$ respectively. Under a further change of
variables, $x=\sig-\sig_c(s)$, $y=\phi-\phi_c(s)$ and $s=u$, the
evolution equation \form{NEV} can be written as
\be\ba{rl}
{\displaystyle
\frac{\partial P_c}{\partial s}\simeq }
&\!{\displaystyle -
\left(\frac{\partial A_\sig}{\partial \sig_c}(\sig_c,\phi_c)\right)
\frac{\partial}{\partial x}[x P_c(x,y)] + B(\sig_c,\phi_c)
\frac{\partial^2 P_c}{\partial x^2} } \\[3mm]
&\!{\displaystyle -
\left(\frac{\partial A_\phi}{\partial \phi_c}(\sig_c,\phi_c)\right)
\frac{\partial}{\partial y}[y P_c(x,y)] + B(\sig_c,\phi_c)
\frac{\partial^2 P_c}{\partial y^2} },
\ea\eel{DPS}
where ${\displaystyle A_\sig(\sig,\phi)-A_\sig(\sig_c,\phi_c)
\simeq x \frac{\partial A_\sig}{\partial \sig_c} +
y \frac{\partial A_\sig}{\partial \phi_c}\ }$ and
${\displaystyle\ A_\phi(\sig,\phi)-A_\phi(\sig_c,\phi_c)
\simeq y \frac{\partial A_\phi}{\partial \phi_c} +
x \frac{\partial A_\phi}{\partial \sig_c} }$. The cross-terms
have been neglected since they contribute with vanishing
correlators.

Under the initial condition $P(x,y;0)=\delta(x-x_o)
\delta(y-y_o)$, the dispersion coefficients can be calculated
as \cite{GonLin}
\be\ba{rl}
{\displaystyle
\Delta^2_\sig = 2 A_\sig^2\ \int_{\sig_o}^{\sig_c} {B\over A_\sig^3}
d\sig }=&\!{\displaystyle
\frac{\lam}{3n^2} \left({\w\over2\pi}\right)^2
\frac{\sig_c^{2n-2}}{\phi_c^4} (\sig_o^4-\sig_c^4)
\sim \frac{\lam}{3n^2} \left({\w\over2\pi}\right)^2
\frac{\sig_c^{2n-2} \sig_o^4}{\phi_c^4} }\\[4mm]
{\displaystyle
\Delta^2_\phi = 2 A_\phi^2\ \int_{\phi_o}^{\phi_c} {B\over A_\phi^3}
d\phi }=&\!{\displaystyle
\frac{\lam}{3n} \left({\w\over2\pi}\right)^2
\sig_c^{2n}\left(\frac{1}{\phi_o^2} - \frac{1}{\phi_c^2}\right)
\sim \frac{\lam}{3n} \left({\w\over2\pi}\right)^2
\frac{\sig_c^{2n}}{\phi_o^2}.}
\ea\eel{DWZ}
The probability distribution $P_c$ then looks like a Gaussian
with decreasing dispersions along both field directions,
centered on the classical field trajectory in the plane
$(\sig,\phi)$,
\be
P_c(\sig,\phi;t) \sim \exp\left\{-\frac{(\sig-\sig_c(t))^2}
{2\Delta^2_\sig(t)}-\frac{(\phi-\phi_c(t))^2}{2\Delta_\phi^2(t)}
\right\}.
\eel{PRB}
This distribution describes the diffusion of the scalar fields
in comoving space after a given initial condition, and shows
how the fields follow the classical trajectory with variable
dispersion. However, in this description we are not taking into
account the expansion of physical space.

\subsection{Self-reproduction of the inflationary universe}

If all the inflationary domains contained fields that followed
their classical trajectory with a given dispersion, all would
have ended inflation by now. However, it was realized in ref.
\cite{SelfRep} that those few domains that jump opposite to the
classical trajectory contribute with a larger physical space and
therefore dominate the physical volume of the universe. Those
domains will split into smaller domains, some with lower values
of the scalar fields, some with higher. As a consequence of the
diffusion process, there will always be domains which are still
inflating, and this corresponds to what is known as the
self-reproduction of the inflationary universe.

We are thus interested not in the probability distribution in
the comoving frame $P_c$, but on the distribution
$P_p(\sig,\phi;t)$ over the physical volume, which takes into
account the different quasi--exponential growth of the proper
volume in different parts of the universe,
\be
P_p(\sig,\phi;t)\simeq P_c(\sig,\phi;t)
\ e^{3\Delta t(H-H_c)}
\eel{PPC}
where $P_c(\sig,\phi;t)$ is given in \form{PRB}. We can now
study the onset of the self--reproduction of the inflationary
universe. The bifurcation line is defined by the values of the
scalar fields for which the maximum of the probability
distribution $P_p$ \form{PPC} starts increasing, or
equivalently, where the quantum diffusion of the fields
dominates its classical motion. Its expression is best found in
polar coordinates, in which the classical motion takes the form,
\be\ba{rl}
{\displaystyle
z \equiv {\phi\over\varphi} }&\!{\displaystyle \longrightarrow
\ {\dot{z}\over z} = {H\over\w} \left(1+z^2\right) },\\[4mm]
r^2 \equiv \phi^2 + \varphi^2 &\!{\displaystyle
\longrightarrow \ \dot{r}=0,}
\ea\eel{ZR}
while the quantum diffusion steps can be written as
\be\ba{rl}
{\displaystyle
{\delta z\over z} = \left(\left({\delta\phi\over\phi}\right)^2 +
\left({\delta\varphi\over\varphi}\right)^2\right)^{1/2} }=
&\!{\displaystyle
{H\over2\pi\phi} \left(1+z^2\right)^{1/2} },\\[3mm]
{\displaystyle
r \delta r = \left(\left(\phi\delta\phi\right)^2 +
\left(\varphi\delta\varphi\right)^2\right)^{1/2} }=
&\!{\displaystyle {H\over2\pi}\ r .}
\ea\eel{DZR}
It is clear that along the radial direction quantum diffusion
will always dominate classical motion, since $\dot r=0$.
However, along the angular direction the condition that quantum
diffusion dominates classical motion during the time interval
$\Delta t = H^{-1}$ ensures the self--reproduction of the
universe only in the range \cite{ExtChaot}
\be
{3\M^4(\phi)\over4\w}\left(1+{\phi^2\over\varphi^2}\right) <
V(\sig) < \M^4(\phi).
\eel{BIF}
Depending on the value of $\w$, this could be a very wide range
of values for the scalar fields. Once an inflationary domain
enters this regime it will create more domains with growing
values of the scalar fields. In particular, the inflaton field
will tend to diffuse towards larger and larger values, which
favor a larger volume of physical space. A reasonable question
would be to ask if there is a stationary solution to the upward
drive of the inflaton due to its quantum diffusion.  This will
be examined in the next section.

\section{\label{stat} Stationary Solutions}

The best way to analyze the possibility of a stationary regime
to the upward diffusion of the inflaton is to look for
stationary solutions of the Fokker--Planck equation for the
probability distribution in the physical frame,
\be\ba{rl}
{\displaystyle
\frac{\partial P_p}{\partial t} }=&{\displaystyle
\frac{\partial}{\partial\sig}\left[\frac{\M^2(\phi)}
{4\pi} \frac{\partial H}{\partial\sig} P_p + \frac{H^{3/2}}
{8\pi^2} \frac{\partial}{\partial\sig}\left(H^{3/2}P_p \right)
\right]} \\[3mm]
+&{\displaystyle
\frac{\partial}{\partial\phi}\left[\frac{\M^2(\phi)}
{2\pi} \frac{\partial H}{\partial\phi} P_p + \frac{H^{3/2}}
{8\pi^2} \frac{\partial}{\partial\phi}\left(H^{3/2}P_p \right)
\right] + 3HP_p. }
\ea\eel{3HP}
Due to the term $3HP_p$ there are no simple stationary solutions.
However, with the use of the identity $D^2g - 2D(Df\ g) = e^f
D^2(e^{-f} g) - [(Df)^2 + D^2f]g$, we can write the stationary
probability distribution for an arbitrary potential as
\be
P_p(\sig,\phi;t) \propto  e^{E t} H(\sig,\phi)^{-3/2}
\exp\left(\frac{3\M^4(\phi)}{16V(\sig)}
\right)\Psi(\sig,\phi),
\eel{PSI}
where the new function $\Psi(\sig,\phi)$ satisfies a
two--dimensional Sch\"odinger--like equation,
\be
\left(H^{3/2}\frac{\partial}{\partial\sig}\right)^2\Psi +
\left(H^{3/2}\frac{\partial}{\partial\phi}\right)^2\Psi -
V(\sig,\phi) \Psi = 8\pi^2 E \Psi,
\eel{FEP}
with an effective potential
\be\ba{rl}
V(\sig,\phi) =&{\displaystyle
\frac{16\pi^4}{9} H^{-5}(\sig,\phi) V'(\sig)^2 +
\frac{4\pi^2}{3} H^{-1}(\sig,\phi) \left(\frac{5}{4}
\frac{V'(\sig)^2}{V(\sig)} - V''(\sig)\right) }\\[3mm]
+&{\displaystyle \frac{64\pi^4}{3\w} H^{-3}(\sig,\phi) V(\sig)
+\frac{6\pi^2}{\w} H(\sig,\phi) - 24\pi^2 H(\sig,\phi) }.
\ea\eel{VSP}
For general theories of the type $V(\sigma) = {\lambda\over2n}
\sigma^{2n}$, the Schr\"odinger potential \form{VSP} can be
written as
\be\ba{rl}
{\displaystyle
\left({2\w\lam\over 3n}\right)^{3/2} V(\sig,\phi) }=&
{\displaystyle \left({2\pi\over\w}\right)^4
\left(\frac{3n^2}{4\lam}\right)^2 {\phi^5\over\sig^{n+2}} +
\left({2\pi\over\w}\right)^2 \frac{3n^2(n+2)}{8\lam}
{\phi\over\sig^{2-n}} }\\[4mm]
+&{\displaystyle \left({2\pi\over\w}\right)^4
\left(\frac{3n}{2\lam}\right)^2 {\phi^3\over\sig^n} +
\left({2\pi\over\w}\right)^2 \left(\frac{9n}{4\lam}\right)
{\sig^n\over\phi} - {2\pi\over\w}
\left(\frac{18\pi n}{\lam}\right) \frac{\sig^n}{\phi}. }
\ea\eel{VXY}

Even in this case it is extremely difficult to find exact
analytical solutions to the Schr\"odinger equation \form{FEP},
due to the non-linear coupling of the two scalar fields.  In
ref. \cite{GBLL} we performed a numerical simulation that gave
us insight into the behavior of the distribution.  We will try
to give here a general idea by analyzing a simple generic case,
$n=2$.  By transforming into polar variables \form{ZR}, we
notice that the Universe will enter the regime of
self-reproduction for $z < (\lambda\w^3/12\pi^2)^{1/4} \ll 1$
and therefore we can approximate the diffusion equation by
\be
\left({r^{3/2}\over z^{3/2}}{\partial\over\partial r}
\right)^2\Psi + \left({r^{1/2}\over z^{3/2}}{\partial\over
\partial z}\right)^2\Psi + {72\pi^2\over\lam\w}{r\over z}
\Psi = 8\pi^2 E \left({3\over\lam\w}\right)^{3/2}\Psi.
\eel{DFF}
The same approximation ($\lam\w^3\ll1$) ensures that the second
term, corresponding to the diffusion in the angular direction,
will dominate over the first. Therefore, it is expected that due
to the potential, the distribution will become concentrated
close to the Planck boundary, where $z\simeq z_p =\sqrt\lam\
\w/4\pi$.  Once there, the distribution will slide in the radial
direction along the Planck boundary, see ref. \cite{GBLL}. Under
the change of variables $r=4/s^2$, the diffusion equation can
then be approximated by
\be
{\partial^2\Psi\over\partial s^2} + {18\w\over s^2}\Psi = E'' \Psi,
\eel{FDF}
where $E''$ includes a term coming from the $z$--diffusion.
Equation \form{FDF} does not have a stationary solution. It is
well know from quantum mechanics that a potential proportional
to $-1/s^2$ is too strong and produces singular solutions at
$s=0$ $(r=\infty)$. The probability distribution will move
forever towards large $r$, {\it i.e.} it is non-stationary. This
result is generic, it is true for all potentials of the type
$\lam\sig^{2n}$.  Those probability distributions were named
``runaway solutions" in ref. \cite{GBLL}.

Runaway solutions are a special feature of Jordan--Brans--Dicke,
since the evolution in general relativity is equivalent to the
motion along the $z$ direction, which is stationary. The
dynamics of the gravitational constant introduces the $r$-motion
along the Planck boundary, that is non-stationary unless we
impose further boundary conditions. For instance, it was shown
numerically in ref. \cite{GBLL} that a sharp increase in the
inflaton potential could effectively simulate a boundary
condition in $\sig$, and therefore produce global stationary
solutions.

\section{\label{conc} Conclusions}

In this paper we have described the mathematical formalism for
stochastic inflation in the context of Jordan--Brans--Dicke
theory of gravity. If string theory is the correct description
for the gravitational interaction at the quantum level, it is
important to know the behavior of the massless gravitational
sector of strings in the low energy effective theory, below the
Planck scale. Such a sector contains a dilaton scalar field as
well as the graviton, and therefore it is worthwhile studying
the cosmology of scalar-tensor theories of gravity. The dilaton
acts like a dynamical gravitational constant, the Brans--Dicke
field. Even though it is argued that the present experimental
observations put severe constraints on such a scalar component
of gravity, we cannot disregard its effects close to the Planck
scale, where quantum fluctuations of the fields become
important.

Since chaotic inflation, it was realized that the inflationary
regime went all the way up to the Planck scale. The quantum
fluctuations of the fields then became important and their
effect on their own background fields was described with the use
of diffusion equations. Stochastic inflation is the mathematical
formalism that describes the Brownian motion of those
coarse-grained background fields during inflation. It is also a
very powerful tool for the description of the very large scale
structure of the Universe, much beyond our observable universe.
Its probabilistic description opened the possibility of a
connection with quantum cosmology \cite{LinBook}.  In order to
incorporate the effect of the string dilaton in the description
of the inflationary universe close to the Planck scale, we have
studied the stochastic inflation formalism in the context of the
simplest scalar-tensor theory, Jordan--Brans--Dicke theory of
gravity.

For that purpose we described the Arnowitt--Deser--Misner
formalism for JBD theory with an inflaton scalar field, in order
to include in a consistent way the non-linear evolution and
backreaction of the scalar fields and the metric. We then
separated the fields into long-wavelength background fields and
short-wavelength quantum fluctuations, thanks to the natural
scale of the de Sitter horizon in inflationary domains, and
obtained a self-consistent set of equations for the background
fields by expanding in spatial gradients. Solutions to these
equations are easily found in the slow-roll approximation,
describing the classical evolution of the scalar fields.

Quantum fluctuations of the scalar fields during inflation give
rise to adiabatic energy density perturbations that could be
responsible for the temperature fluctuations recently observed
by COBE in the cosmic background radiation. We describe a
consistent way of computing the amplitude of the perturbations
on a homogeneous background in terms of the quantum fluctuations
of both dilaton and inflaton scalar fields. For theories of the
type $\lambda\sig^4$ we find the usual scale-invariant
$\sqrt\lam$ dependence. However, for theories with a mass term
for the inflaton, the amplitude of adiabatic density
perturbations is proportional to the ratio of such a fundamental
scale (assumed given) and the effective value of the Planck
scale at the end of inflation.

We then studied the new stochastic phenomena that appear when
the dilaton is included in the diffusion equations for the
probability distribution of finding a certain value of the
scalar fields both in the comoving and the physical frame. While
the probability distribution in the comoving frame followed the
classical trajectories of the scalar fields, with some
time-dependent dispersion, the description in the physical frame
takes into account the quasi-exponential growth of the
inflationary domains. Those few domains that jump opposite to
the classical trajectory inflate more and thus contribute with a
larger physical space. There is a bifurcation point (in fact a
line in JBD inflation) at which the contribution to the proper
volume of the Universe of a given inflationary domain dominates
the tendency of the fields to follow their classical
trajectories, and the Universe enters a regime of
self-reproduction.  In that regime some domains end inflation
and enter the radiation and matter dominated eras, while others
continue inflating and producing new inflationary domains,
overwhelmingly dominating the total proper volume of the
Universe. In that sense, our own observable universe might be an
insignificant subproduct of one of those domains that ended
inflation \cite{LLM}. We compute the bifurcation line in the case
of JBD stochastic inflation and find that the range of values of
the scalar fields in the self-reproduction regime depends on the
value of the Brans--Dicke parameter.

The main question to be addressed is the existence of stationary
solutions for the probability distribution in the physical
frame. With the introduction of the dilaton field, the diffusion
space is now bidimensional and the self-reproduction of the
Universe occurs on an infinite wedge in this space, see eq.
\form{BIF}, bounded between the bifurcation line and the Planck
boundary although open for arbitrary large values of the scalar
fields. We have found, see also ref. \cite{GBLL} for a more
detailed description based on numerical simulations, that in
general we will obtain runaway solutions: non-stationary
probability distributions concentrated close to the Planck
boundary but moving with different speeds towards large values
of the scalar fields, in particular to an infinitely large
Planck mass. Perhaps this is an indication that we need to
introduce a dynamical cutoff in the inflaton potential. In ref.
\cite{GBLL} we discussed the cosmological consequences of this
result.

In the case of general relativity, the energy eigenvalue of the
stationary distributions was related to the fractal dimension of
the global inflationary universe \cite{Fractal}.  In our case,
the existence of runaway solutions indicate a very complicated
fractal structure of the universe close to the Planck scale.
Further work should be done in this direction.

\section*{Acknowledgements}

It is a pleasure to thank Andrei Linde for lively discussions
and Arthur Mezhlumian for a careful reading of the manuscript.
Work supported by a Spanish Government MEC-FPI Postdoctoral
Grant.

\end{document}